\documentstyle[prl,aps]{revtex}
\begin{document}
\title{Pathways in Two-State Protein Folding}
\author{Audun Bakk and Johan S.\ H{\o}ye}
\address{Department of Physics, Norwegian University of Science and 
Technology, NTNU, N--7034 Trondheim, Norway}
\author{Alex Hansen\footnote{Permanent Address: Department of Physics, 
Norwegian University of Science and Technology, NTNU, N--7491 
Trondheim, Norway} and Kim Sneppen\footnote{Permanent Address:
NORDITA, Blegdamsvej 17, DK-2100 {\O}, Denmark}}
\address{International Centre for Condensed Matter Physics,
University of Bras{\'\i}lia, CP 04513, 70919--970 Bras{\'\i}lia--DF,
Brazil}
\author{Mogens H.\ Jensen}
\address{Niels Bohr Institute, Blegdamsvej 17, DK-2100 {\O}, Denmark}
\date{\today}
\maketitle
\begin{abstract} 
The thermodynamics of proteins indicate that folding/unfolding
takes place either through stable intermediates or through a 
two-state process without intermediates. The rather short
folding times of the two-state process indicate that folding is guided. 
We reconcile these two seemingly contradictory observations 
quantitatively in a schematic model of protein folding. 
We propose a new dynamical transition temperature which is lower than 
the thermodynamic one, in qualitative agreement with 
in vivo measurement of protein stability using {\it E.coli}.
Finally we demonstrate that our framework is easily
generalized to encompass cold unfolding, and make predictions
that relate the sharpness of the cold and hot unfolding 
transitions.
\vspace{0.5cm}

\end{abstract} 
\vskip0.5cm
Proteins fold to a uniquely defined ground state,
and do this in spite of the astronomical number
of possible states. This paradox, usually attributed to
Levinthal, is further sharpened
in view of the fact that 
there is thermodynamic evidence 
for the folding transition behaving nearly
as for a two-state system for a large 
sub-class of single domain proteins 
\cite{pk74,p92,br99a,br99b}.
One would think that an on-off
process would exclude the possibility 
of guiding, and indeed simple
guiding predicts a first order phase transition
which is far softer than experimentally found for the two-state 
class of proteins.   
The purpose of this letter is to quantify the degree of
guiding that is compatible with the observed two-state folding
process.  We do this through generalizing a hierarchical 
protein model introduced earlier in Ref.\ \cite{hjsz98b}. 
In this model guiding dominates the dynamics of 
the folding process, 
which in this frame is defined through the Monte Carlo (MC) procedure
applied when simulating the stochastic behavior of the model.  
We find here that guiding can dominate the dynamics of folding 
and still maintain the thermodynamic behaviour as that of a 
two-state process.

Another prediction of our scenario is that the cold unfolding transition 
\cite{cs89,p92} should exhibit a sharpness close to that of the
hot unfolding transition. To our knowledge there is no 
experimental studies of the sharpness of the cold unfolding transition.

The van't Hoff relation \cite{p92}, 
\begin{equation}
\label{vanthoff}
\Delta H=\alpha k T_c^2 \frac{\Delta C}{Q}\;,
\end{equation}
provides a powerful way to quantify the sharpness 
of a first order phase transition taking place at $T_c$.  
It relates the enthalpy difference
between the two phases, $\Delta H$ to the height of the 
heat capacity peak, $\Delta C$ and latent heat of the
transition $Q$ with a proportionality factor $\alpha$.
A smaller $\alpha$ corresponds to a sharper transition.
When the transition is two state, $\alpha =4$, and when the transition has
a large number of equally stable intermediates, $\alpha=12$.
For the single domain proteins, ribonuclease, lysozyme, chymotrypsin,
cytochrome c and myoglobin, Privalov and Khechinasvili \cite{pk74} find
experimentally
\begin{equation}
\label{privalov}
\alpha = 4.2
\end{equation}
to within 5 \% accuracy,
demonstrating that these transitions are very nearly two-state.
Other small proteins like $\alpha$-lactalbumin,  RNase H,
barnase and cyt c have known metastable intermediates \cite{br99a}. 

The two extremes of protein folding is spanned by respectively
the two-state and a multiple state ``zipper''-like 
\cite{s58,dbyfytc95}
description of the process \cite{hjsz98b}. 
We sketch the model and its parametrization briefly here.
The relevant degrees of freedom (conformational angles) are 
modeled through binary variables $\psi_i$.  
They either locally match the 
ordered structure: $\psi_i=1$, or they do not: $\psi_i=0$.  
Guiding is imposed through the series of inequalities
\begin{equation}
\label{constraint}
\psi_i \ge \psi_{i+1}\;.
\end{equation}
The variables $\psi_i$ alone 
cannot describe the degrees of freedom that become liberated when a portion
of the protein is not matching the local structure of the native state. 
In order to take these extra degrees of freedom into account,  
a second, independent set of variables, $\xi_i$, is introduced. 
For simplicity, these variables are assigned the values 1 or $1-E$.  
The Hamiltonian is
\begin{equation}
\label{psiham}
H~=~-\sum_{i=1}^N \psi_i\xi_i\;,
\end{equation}
with the constraints (\ref{constraint}) in effect.  The interpretation of  
the terms in this Hamiltonian is that when there is a local match, 
$\psi_i=1$, there is a energy cost of $E$ to change the $\xi_i$ variable.
When there is no match, there is no energy cost associated with
changing $\xi_i$ --- it ``flaps" freely.

We note that for any finite value of $E$, the protein may change structure
locally due to change in $\xi_i$
even in the parts of the protein where $\psi_i=1$.
In order to simplify the analysis, we assume $E$ to be sufficiently large
compared to any other energy scale in the system --- in particular $kT$,
where $T$ is the temperature --- so that the $\xi_i$ variables never
take the value $1-E$ when $\psi_i=1$.

We may define a set of binary, unconstrained
variables $\varphi_i$, taking the values zero or one such that
\begin{equation}
\label{psiphi1}
\psi_i=\varphi_1\cdots\varphi_i\;.
\end{equation}
In particular, $\psi_1=\varphi_1$.  In the limit when $E \to\infty$, 
the Hamiltonian (\ref{psiham}) becomes
\begin{equation}
\label{phiham}
H_{p1}~=~-\varphi_1-\varphi_1\varphi_2-
\varphi_1\varphi_2\varphi_3+\cdots -
\varphi_1\varphi_2\cdots\varphi_N\;,
\end{equation}
where there are no additional constraints.  
The role of the variables $\xi_i$ is now played by the 
degeneracy present in (\ref{phiham}).

One can easily show that (\ref{phiham}) has a first order
phase transition at $T_c=1/\log(2)$ where the ordered phase 
$\{\varphi_i\}=\{1111\cdots 1\}$ with energy $U=-N$
melts to a disordered structure with energy $U\approx 0$.
Thus $\Delta H=Q=N$ and $\Delta C = N^2 \log(2)^2/12$
leading to $\alpha=12$. On the other hand
if we only consider a rescaled last term\
\begin{equation}
\label{phiham4}
H_{p2}= -N \varphi_1\varphi_2\cdots\varphi_N\;,
\end{equation}
then one also obtains a phase transition at $T_c=1/\log(2)$, with
$\Delta H=Q=N$ but with $\Delta C= N^2 \log(2)^2/4$. 
Thus in this case $\alpha=4$.
There is no guiding in the Hamiltonian (\ref{phiham4}) since the ground
state, $\{1111\cdots 111\}$, is one out of the $2^N$ possible states, while 
all the other $2^N-1$ states are degenerate.  

We define time in the model based on the MC method \cite{b87}.  
The values of $\phi_i$
are chosen or changed randomly, and acceptance of each choice depends upon
the usual Boltzmann factor due to any energy shift connected to this.  Time
advances by one unit for every attempted update of the $\phi_i$ variables.
We note, however, that the dynamics of an MC procedure may be different 
from the actual dynamics of a given Hamiltonian, although properties at 
thermal equilibrium are properly represented.  

The average folding time 
measured as the typical number of states visited before finding the 
ground state is widely different in the two models.
For the true two-state model (\ref{phiham4}) the average folding time is
$2^N/2$.  For the guided system governed by (\ref{phiham}) 
the ground state is found in a time growing as $N^2$ when $T$ is below
$T_c$.  To reconcile that a large class of proteins   
behave as a two-state system with the necessity of being able to reach 
the ground state in a reasonable time, we now study a combination of the two 
Hamiltonians (\ref{phiham}) and (\ref{phiham4})
\begin{equation}
\label{phiham5}
H_p \; =\; \lambda_p H_{p1}\; +\; (1-\lambda_p) H_{p2}\;.
\end{equation}
This Hamiltonian has a transition at $T_c=1/\log(2)$ for all
values of $\lambda_p$. The behaviour can be parametrized 
by the smallest $n$ for which $\varphi_{n+1}=0$. 
For a given temperature the partial free energy of states 
characterized by $n$ is 
$F(n)\; =\; n \; ( T \log(2) - \lambda_p ) \; -\; 
\delta_{N,n} \; ( 1 - \lambda_p ) \; N$.
In Fig.\ \ref{fig0} we show $F(n)$ schematically for different temperatures $T$.
Each $F(n)$ exhibits a jump at $n=N$ corresponding to the 
energy gain $N (1-\lambda_p)$ for reaching the ground state.
At low $T$, $F(n)$ is monotonically decreasing, reflecting a fast folding
kinetics where the typical folding time grows as $N^2$. 
At an intermediate $T=T_G=\lambda_p/\log(2)$ all $n<N$
are equally probable. For $T$ in the interval between $T_G$
and $T_c$ the intermediate states are unstable 
(see Fig.\ \ref{fig0} ) ---
i.e.\ they form a barrier between the folded and denatured state ---
and the folding time scale exponentially with both $T$ and $N$.
At a higher $T=T_c=1/\log(2)$
the folded state becomes unstable, and the protein melts 
($\langle n\rangle \approx 0$).
The fact that the free energy landscape changes with $T$
means that two-state folding around $T_c$ is compatible
with guiding and fast folding at low $T$.

Fig.\ \ref{fig1} shows the van't Hoff coefficient $\alpha$ as a function
of $\lambda_p$ on the unit interval based on direct calculation of the
partition function.
One observes that increasing $\lambda_p$ --- i.e.\ increasing the guiding 
--- leads to increasing $\alpha$ and thus a softening of the transition.
As $N$ is increased, the regime where $\alpha$ is very close to 4 is
expanded towards higher values of $\lambda_p$.  For example, with the 
experimental observation of $\alpha=4.2$, and assuming $N=10$, 
$\lambda_p$ is close to zero while for $N=100$, $\lambda_p$ is approximately
0.7.  Thus, in this latter case, 70\% of the energy difference between the 
unfolded and folded states sits in the guiding, and still $\alpha$ is very
close to the value indicating the folding process to be essentially
a two-state process. 

We now discuss the fact that large $N$ allows for more 
guiding without destroying the two-state nature of the transition.
To understand this we note that any $\lambda_p<1$ 
in fact define a virtual phase transition
at $T=T_G<T_c$.
At $T_G$ the protein would melt
if it were not due to the additional gain in binding
energy when the ground state is reached. This virtual transition
is not seen directly in equillibrium thermodynamics, 
but strongly influences the dynamic behaviour
in the temperature range between $T_G$ and $T_c$:
In this intermediate regime the protein is a two-state system
where occasionally melting implies a long waiting time
with many partial refreeze attempts.
Due to fluctuations a system with small $N$ 
can be partly refrozen
also above the virtual transition, 
and thus $\alpha$ depends on system size as shown in fig.\ \ref{fig1}.

Experimentally, if one are dependent on dynamics one presumably
measure $T_G$ as the transition temperature, while for experiments
based on thermodynamics it would be $T_c$.
For fast living organisms such as {\it E.coli\/} the overall status
of fraction of unfolded proteins can be monitored by 
the level of chaperone DnaK \cite{Gross,Bourke}.
For temperatures between 13 and 37 C the DnaK 
per {\it E.coli\/} cell raises slowly from 4000 to
6000, whereafter it raises sharply to 
$\sim 8500$ at 42 C and $\sim 18000$ at 46 C
\cite{Herendeen,Pedersen}.
At 50 C the {\sl E.coli} dies.
This may be taken as an indication that in the temperature
interval above 37 C the typical proteins need help
in the folding process. But as the cell is able to sustain life up to 
about 50 C, the typical proteins must have some stability 
up to this higher temperature.
This is reminiscent of the behaviour of our model, 
with a $T_G$ of about 37 C,
an exponentially slow folding of proteins, necessitating the help of 
chaperones, 
for higher temperatures and a $T_c$ of the order of 50 C \cite{hjps}.

The above considerations can be extended to include a more
realistic scenario where the protein is reacting with water.
Following ref.\ \cite{hjsz98b} we parametrize this through water variables 
$w_1, w_2, ... , w_N$, taking values 
${\cal E}_{\min} + s \Delta$, $s=0,1,...,g-1$. Here, $\Delta$ is
the spacing of the energy levels of the water-protein interactions.
We quantify the coupling to the water by a combination of the Hamiltonians
\begin{equation}
\label{eq1}
H_{w1}~=~
(1-\varphi_1) w_1 + (1-\varphi_1\varphi_2) w_2 + ... + 
(1-\varphi_1\varphi_2\cdots\varphi_N) w_N \;,
\end{equation}
and
\begin{equation}
\label{eq2}
H_{w2}~=~
(1-\varphi_1\varphi_2\cdots\varphi_N) (w_1+\cdots +w_N) \;,
\end{equation}
to form the total Hamiltonian 
\begin{equation}
\label{fullham}
H=\lambda_p H_{p1} + (1-\lambda_p) H_{p2}
+ \lambda_w H_{w1} + (1-\lambda_w) H_{w2}\;.
\end{equation}
(Here it may be noted that $H_{w2}$ may introduce {\it non-local\/} 
interactions between distant units, when the terms are interpreted using
the variables $\psi_i$ and $\xi_i$.)
When $\lambda_p=\lambda_w=1$ we are back to the
Hamiltonian defined in ref. \cite{hjsz98b} whereas
when $\lambda_p=\lambda_w=0$ we are facing a 
two-state Hamiltonian.  In Fig.\ \ref{fig2}
we display the heat capacity curves for these two extremes.
The system is folded in its ground state between the 
cold unfolding transition at $T=1.2$ and the hot unfolding
transition at $T=4.7$
As also quantified by the van't Hoff coefficients,
we see that the Hamiltonian without guiding gives
a phase transition which is about a factor 3 sharper
for both the cold and the hot unfolding transitions.
Also in terms of temperature, these transitions are much more 
separated than in real systems where the freezing of water will look
much more like ``absolute zero".  The present model as it stands is not
able to account for this.

In Fig.\ \ref{fig3} we investigate systematically the
van't Hoff coefficient $\alpha$ as function of
$\lambda_p$ and $\lambda_w$ for the hot
(Fig.\ \ref{fig3} a) and the cold (Fig.\ \ref{fig3} b) transition.
As is evident, $\alpha$ is similar but somewhat larger
for the hot than for the cold transition.
As a consequence, the cold transition transition is slightly sharper.
We are not aware of any experimental measurements of the van't Hoff
coefficient for the cold transition. Such a measurement will, however,
in practice be hampered as the cold
transition is mainly seen experimentally at pH-values 
where it is close to the hot transition.

Finally, we note the distinct feature of the
cold transition $\alpha$ when
$(\lambda_p, \lambda_w) \approx  (1,0)$
where it drops to a value below 4.
We will show in \cite{bhhsj99} that this is a remnant 
of the phase transition that would have appeared 
at $T=T_c=1/\log(2)$
if the system were not coupled to the water.

We summarize by noting that in this protein model, it is easy to 
reconcile the thermodynamics of a two-state system 
with the dynamics of a guided system, as this can be done
by diminishing either $\lambda_p$ and/or $\lambda_w$ from
the value one. The dynamical consequence of the hereby masked
guiding is a folding times that is dramatically reduced
when temperature moving away from the transition temperature.

We note as final consequence of our model that good folders can be 
viewed as random sequences of folding steps of which
the last have a particularly favorable binding energy thereby
securing two state cooperativity.
\vspace{0.5cm}

A.B.\ thanks the Norwegian Research Council for financial support.  
A.\ H.\ and K.\ S.\ thank F.A.\ Oliveira and H.N.\ Nazareno for warm
hospitality and the I.C.C.M.P.\ for support during our stay in Brazil.  
We thank G.\ Zocchi for countless discussions.

\begin{figure}
\caption{A schematical drawing of the partial free energy $F(n)$ as function of
the level of folding $n$ for for different temperatures $T$.}
\label{fig0}
\end{figure}
\begin{figure}
\caption{The van't Hoff coefficient $\alpha$ as a function of $\lambda_p$ 
for $N=10$ and 100.}
\label{fig1}
\end{figure}
\begin{figure}
\caption{Heat capacity curves for $N=50$ system 
with and without guiding, i.e. with
$\lambda_p=\lambda_w=1$ respectively
$\lambda_p=\lambda_w=0$. The parameters for the
water variables are $\epsilon=-3.1$, $\Delta =0.04$
and $g=350$.  
}
\label{fig2}
\end{figure}
\begin{figure}
\caption{van't Hoff coefficient $\alpha$ for a) hot respectively b) 
cold transition for $N=100$ system.  Other parameters are as in  
Fig.\ 2.}
\label{fig3}
\end{figure}

\end{document}